\documentclass{IEEEtran}

\makeatletter
\def\endthebibliography{%
	\def\@noitemerr{\@latex@warning{Empty `thebibliography' environment}}%
	\endlist
}
\makeatother

\usepackage{cite}

\usepackage[pdftex]{graphicx} 

\usepackage[caption=false,font=footnotesize]{subfig}

\usepackage{amsmath}
\usepackage{amsfonts}
\usepackage{bm}
\usepackage{filecontents}
\usepackage{amssymb}
\usepackage{color}

\usepackage{epsfig}

\usepackage{verbatim}

\usepackage{url}

\usepackage{algorithm, tabularx}

\usepackage{lettrine}

\usepackage{lipsum}

\usepackage{siunitx}

\usepackage{soul,color}

\usepackage{mathrsfs}

\usepackage{array}

\usepackage[inline]{enumitem}

\usepackage{flushend}

\usepackage{dblfloatfix} 

\usepackage{algorithm}

\usepackage[noend]{algpseudocode}

\usepackage{setspace}
\usepackage{multirow}
\usepackage{array}
\newcolumntype{L}[1]{>{\raggedright\let\newline\\\arraybackslash\hspace{0pt}}m{#1}}
\newcolumntype{C}[1]{>{\centering\let\newline\\\arraybackslash\hspace{0pt}}m{#1}}
\newcolumntype{R}[1]{>{\raggedleft\let\newline\\\arraybackslash\hspace{0pt}}m{#1}}

\DeclareMathOperator{\sinc}{sinc}
\DeclareMathOperator{\argmax}{argmax}
\newcommand{\norm}[1]{\left\lVert#1\right\rVert}

\begin{document}
	
	\title{\LARGE RF Lens Antenna Array-Based One-Shot Coarse Pointing for \\Hybrid RF/FSO Communications}

	%Outage Probability of One-Shot Coarse Pointing with \\RF Lens Antenna Array for Free Space Optics
	
	\author{Hyung-Joo Moon,~\IEEEmembership{Student Member,~IEEE}, Hong-Bae Jeon,~\IEEEmembership{Student Member,~IEEE}, and \\Chan-Byoung Chae,~\IEEEmembership{Fellow,~IEEE}
		\thanks{This work was supported in part by IITP grant funded by the Korea government (MSIT) (No. 2019-0-00685, Free space optical communication based vertical mobile network), and (No. 2016-0-00208, High Accurate Positioning Enabled MIMO Transmission and Network Technologies for Next 5G-V2X Services). (\textit{Corresponding author: C.-B. Chae.})}
		\thanks{H.-J. Moon, H.-B. Jeon and C.-B. Chae are with the School of Integrated Technology, Yonsei University, Seoul 120-749, Korea (e-mail: \{moonhj, hongbae08, cbchae\}@yonsei.ac.kr).}% <-this % stops a space
	}
	
	%\markboth{IEEE Wireless Communications Letters,~Vol.~XX, No.~XX, XXX~2020}
	{}
	%{Shell \MakeLowercase{\textit{et al.}}: Bare Demo of IEEEtran.cls for Journals}
	
	%This work was supported in part by the Institute for Information \& communications Technology Promotion (IITP) grant funded by the Korea government (MSIP) under Grant 1711094030, and in part by the. (\textit{Corresponding author: Chan-Byoung Chae.

	\maketitle
	
	\begin{abstract}

%\textcolor{red}{In upcoming beyond fifth-generation (B5G) communication era, the transmission strategies of free-space optical (FSO) signal are arising as a significant feature and attempted to be applied in various B5G scenarios.} The high directivity of FSO signal allows significantly higher data rate despite of the longer propagation distance. That being said, the major difficulty of the system is to acquire and maintain the alignment between two terminals.
%The main assumption of this characteristic is that both sides should be well aligned to each other. Thus, the accurate link aligning process, i.e., pointing, acquisition, and tracking (PAT) system is taking a significant part in FSO communications.

Because of its high directivity, free-space optical (FSO) communication offers a number of advantages. It can, however, give rise to major system difficulties concerning alignment between two terminals. During the link-acquisition step (a.k.a. coarse pointing), a ground station can be prevented from acquiring optical links due to pointing errors and insufficient information about unmanned aerial vehicle locations. We propose, in this letter, a radio-frequency (RF) lens antenna array to increase the performance of coarse pointing in hybrid RF/FSO communications.
%We demonstrate that a lens antenna array installed on a ground station can significantly improve the accuracy of estimation in the initial step of link acquisition. Furthermore,
The proposed algorithm using a novel closed-form angle estimator, compared to conventional methods, reduces the minimum outage probability by over a thousand times.

%We were able to reduce the uncertainty of location information from one terminal to another by over 60\% within 10km distance by utilizing RF link to assist FSO link connection. An angle of incoming RF beacon signal sent by the other side is estimated by lens antenna array at the ground station, which then combined with GPS information to higher the accuracy of the result. The formulation of an outage probability minimization problem allows determining the practical gain of our angle estimator. We showed that our algorithm considerably reduces outage probability during the link connection and varies the optimal beam divergence of an optical beam.%
		
	\end{abstract}


	\IEEEpeerreviewmaketitle

	\section{Introduction}
%Free Space optical communication %

\IEEEPARstart{R}{esearchers} are developing a major future communication technology known as free-space optical (FSO) communications. FSO has two primary advantages--high capacity and unlicensed frequency bands~\cite{2017cst}. Currently, the tradeoff is twofold--the unpredictability of channels and significant pointing difficulties--making such systems less stable than conventional radio-frequency (RF) communications systems. Therefore, researchers have begun to investigate the use of RF links to compensate for system instability (i.e., hybrid RF/FSO communication systems).
In~\cite{2015wircom, 2010twc}, the authors claimed that single-hop parallel RF/FSO communication is a promising technology that utilizes both robustness and high data rate, the complementary advantages of RF and FSO link.
%Since RF link brings robustness and FSO link brings high data rate into the combined system, the authors in~\cite{2015wircom, 2010twc} describe that single-hop parallel RF/FSO communications are a promising technology.

\begin{figure}[t]
	\begin{center}
		{\includegraphics[width=1\columnwidth,keepaspectratio]
			{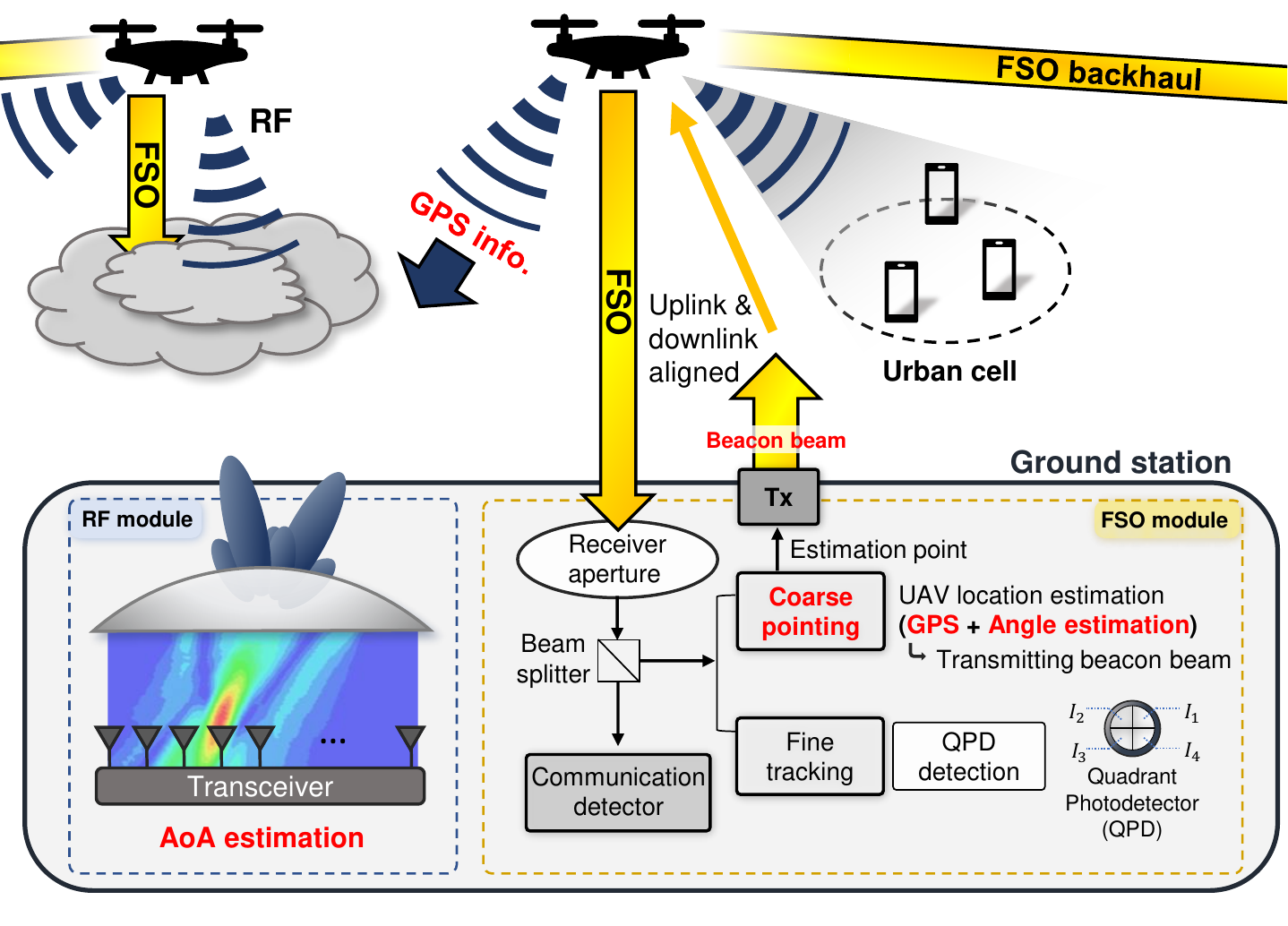}%
			\caption{System model for lens antenna array-based RF/FSO hybrid communications. The key operations of our algorithm are highlighted in red.}
			\label{newfigure}
		}
	\end{center}
	\vspace{-10pt}
\end{figure}

It is widely known that although RF communications have a low risk of link failure, beam-directivity lags far behind that of optical communications. Hence, many studies on hybrid RF/FSO systems assume a link distance of no greater than a few kilometers to maximize the advantages of additional RF links~\cite{2016pj}. One of the approaches in this field is to distribute the high demand for capacitance by deploying unmanned aerial vehicles (UAVs) in an FSO backhaul network; this supplies a sufficient data rate with low latency~\cite{adhocfso}. The UAVs in Fig.~\ref{newfigure}, for example, can be deployed efficiently based on the demands of local users. In this letter, we propose a performance-enhancement strategy for ground-to-air hybrid RF/FSO communications by focusing on link-acquisition capability.

%A common approach is to distribute the high demand for capacitance by deploying unmanned aerial vehicles (UAVs) in an FSO backhaul network to supply a sufficient data rate with low latency~\cite{adhocfso}. For example, the UAVs in Fig.~\ref{newfigure} can be deployed efficiently based on the demands of local users. Other researchers have solved the user scheduling problem of the satellite-UAV-terrestrial network system model based on different criteria~\cite{202008wcl}. In this letter, we propose a performance enhancement strategy for ground-to-air hybrid RF/FSO communications by focusing on link acquisition capability.

%Coarse pointing%

Link-acquisition capability is determined by the performance of the pointing, acquisition, and tracking (PAT) system and is generally broken down into two steps--coarse pointing and fine tracking. The coarse-pointing process begins with the transmission of an optical beacon signal from a ground station to the location that is most likely to be occupied by a UAV. In most hybrid RF/FSO communication systems for ground-to-air applications, this location is determined by the ground station using global positioning system (GPS) information sent from a UAV through RF signals~\cite{2006spienews}. We propose utilizing RF signals not only for backup data links and passing GPS information, but also for angle of arrival (AoA) estimation at ground stations to reduce UAV location uncertainty and provide a reliable link-formation process. When a UAV successfully detects a beacon sent from a ground station, it transmits an optical beam back in the direction of arrival using a transmitter that is tightly aligned with its receiver. Once the ground station receives a beam sent back from a UAV, both the ground station and UAV begin transmitting to each other and controlling the beam direction using fine-tracking modules~\cite{2017cst}. 

%As shown in Fig.~\ref{newfigure}, fine tracking typically utilizes a quadratic photodetector for misalignment detection. It detects biases in the $x$ and $y$ axes independently by calculating differences between the photocurrents generated from each quadrant~\cite{2017cst}.

%~\cite{2020sensors}

%lens antenna%

From the perspective of a PAT system, an RF lens antenna array is an appropriate choice for hybrid RF/FSO systems for several reasons. First, prior work has demonstrated that the Cramér-Rao lower bound of AoA estimation with a lens antenna array is considerably lower than that with a uniform linear antenna array~\cite{awpllens}.
Moreover, since most of the received power is concentrated around the focal point, activating only a few antenna elements does not bring much performance degradation but allows a low-complexity RF beam-tracking on multiple UAVs due to the reduced RF chains~\cite{2016tmtt, 2018commag, 2020wcnc}.
In addition, even if the UAVs use the same bandwidth, signal interference may be resolved by antenna selection of the lens antenna if they are in different directions.
This antenna selection property is also considered in this letter to reduce the complexity of our algorithm. Finally, RF signal-based UAV tracking is possible~\cite{2014ssc}. This allows a ground station to select the most appropriate UAV for communication depending on the channel status and quickly establish an optical link through the algorithm we will propose.
%Following the link-acquisition, a fine tracking control loop maintains the link alignment between two terminals and,
As shown in Fig.~\ref{newfigure}, an RF module receives RF signals from multiple UAVs. When link failure occurs, this system facilitates rapid reacquisition of an FSO link. 

%From the aspect of PAT system, an RF lens antenna array will be an appropriate choice in hybrid RF/FSO systems for several reasons. First of all, the previous research asserts that the Cramér-Rao lower bound of the AoA estimation with lens antenna array is much lower than that with the uniform linear antenna array~\cite{awpllens}. Moreover, due to the RF switch-based low-complexity beamforming property of the lens antenna array structure, continuous tracking and control on multiple UAVs are possible~\cite{2016tmtt, 2018commag}. An RF signal-based UAV tracking is similar to the closed-loop coarse control in satellite FSO communication~\cite{2014ssc}. This allows the ground station to select the most appropriate UAV to communicate depending on the channel status and quickly build the optical link. After the link acquisition, a fine tracking control loop maintains the link alignment between two terminals, while an RF module receives RF signals from multiple UAVs, as depicted in Fig.~\ref{newfigure}. It allows quick re-acquisition of the FSO link when link failure has occurred.

The rest of this letter is organized as follows. In Section~\ref{systemmodel}, we introduce the lens antenna array structure and optical channel model used in our analysis. In Section~\ref{closedform}, we present a closed-form expression of this estimator for a given GPS information. In Section~\ref{algorithm}, a novel one-shot, coarse-pointing algorithm is proposed and described from the perspective of outage probability and processing time. Also proposed in Section~\ref{algorithm} is an evaluation method for the processing time of one-shot, coarse-pointing. Numerical simulation results are presented in Section~\ref{numres} and in Section~\ref{conclusion} we offer our conclusions.

%In Section~\ref{systemmodel}, we first introduce the lens antenna array structure and optical channel model used for analysis. We then derive an AoA estimator with low complexity in the lens antenna array-based hybrid RF/FSO systems depicted in Fig.~\ref{newfigure}. The closed-form expression of the estimator under given GPS information will be proposed in Section~\ref{closedform}. In Section~\ref{algorithm}, a novel one-shot coarse pointing algorithm is newly suggested and described in the perspective of outage probability and processing time. The evaluation method of the processing time of one-shot coarse pointing will be shown. The numerical simulation results are provided in Section~\ref{numres}, and then we conclude in Section~\ref{conclusion}.

\section{System Model}
\label{systemmodel}

Fig.~\ref{newfigure} presents an integrated scenario with an RF lens antenna array and optical PAT module for a hybrid RF/FSO system. The RF module at the ground station utilizes a lens antenna array to track multiple UAVs. It leads to a fast acquisition of an optical link when the high-throughput backhaul or feeder link is required at a specific aerial node.
%To provide an FSO backhaul network, engineers can deploy UAVs as mobile base stations in dense areas or, in sparse areas, as relaying nodes.

\subsection{Signal Model for RF Lens Antenna Array}

The received power of incoming RF signals through a lens differs for each antenna element at the focal distance. The received signal vector $\bold{x}$ can be modeled as ${\bold{x} = g\bold{H}(\phi)\bold{r}e^{jb}+\bold{z}}$, where the diagonal amplitude matrix $\mathbf{H}(\phi)$ is given as follows~\cite{awpllens}:
\begin{align}
\label{ampfunc}
\bold{H}(\phi)_{n+\frac{N+1}{2},n+\frac{N+1}{2}} = \frac{L}{\sqrt{z}} \sinc{\left[\frac{L}{\lambda}\left(\frac{dn}{z}-\sin(\phi)\right)\right]},
\end{align}
where the index $n$ of each antenna element is a set from $-(N-1)/2$ to $(N-1)/2$, and $N$ is the number of antenna elements. The vector $\mathbf{r}$ of length $N$, which varies with the shape of the antenna array, has the form of
\begingroup
\small
\begin{equation}
\label{steering}
\bold{r} = \begin{cases}
\begin{aligned}[b]
s\Bigg[\exp\Bigg\{-\frac{j2\pi}{\lambda} \sqrt{d^2 \left(-\frac{N-1}{2}\right)^2 + f^2}\Bigg\}, \cdots
\\, \exp\Bigg\{-\frac{j2\pi}{\lambda} \sqrt{d^2 \left(\frac{N-1}{2}\right)^2 + f^2}\Bigg\}\Bigg] 
\end{aligned}	&\text{(linear)}\\
s\left[1,\cdots,1\right]^{T}\exp\left\{{-2\pi z/\lambda}\right\} &\text{(arc)},
\end{cases}
\end{equation}
\endgroup
where $s$ is a unit-power signal. The symbols $g$, $b$, ${\phi}$, and $\mathbf{\bold{z}}$ are the signal gain, signal phase, AoA of the signal, and zero-mean complex normal vector with a standard deviation of ${\sigma_\text{n}}$, respectively. The parameters ${L}$, ${d}$, ${n}$, ${\lambda}$ and ${z}$ denote the lens diameter, antenna spacing, antenna index, signal wavelength, and distance between the center of the lens and each antenna, respectively. The value of ${z}$ is determined as ${z=f}$ for an arc antenna array, and ${z=\sqrt{d^2n^2+f^2}}$ for a linear array.

Compared to a uniform linear antenna array, a lens antenna array has the advantage of requiring a small number of RF chains. Suppose the receiver only has $k$ RF chains. At this point, the natural option is to estimate the focal point by substituting an angle $\phi_\text{gps}$ calculated by GPS information into (\ref{ampfunc}), and select the nearest $k$ antennas. The smallest index $u$ and the largest index $v$ can be expressed as ${u=[\frac{z}{d}\sin(\phi_{\text{gps}})-\frac{k}{2}+\frac{1+(-1)^N}{4}]+\frac{1-(-1)^N}{4}}$ and ${v=[\frac{z}{d}\sin(\phi_{\text{gps}})+\frac{k}{2}+\frac{1+(-1)^N}{4}]+\frac{1-(-1)^N}{4}}$. Therefore, the received signal vector with the limited number of RF chains becomes ${\bold{y} = g\bold{A}(\phi)\bold{s}e^{jb}+\bold{n}}$, where $\bold{y}=\bold{x}[u:v]$, $\bold{A}(\phi)=\bold{H}(\phi)[u:v,u:v]$, $\bold{s}=\bold{r}[u:v]$, and  $\bold{n}=\bold{z}[u:v]$.

\subsection{Optical Channels for Performance Evaluation}
\label{chamod}

A received power ${P_{R}}$ can be modeled as ${P_{R} = {h_{\ell}}{h_{a}}{h_{p}}RP_{T}}$, where each of the parameters indicates attenuation, fading, pointing loss, responsivity, and transmitted power, respectively~\cite{2016ptl}.

%%%%%%subsubsection
\subsubsection{Attenuation Loss}
The term ${h_\ell}$ representing attenuation loss follows the Beer-Lambert law as
\begin{equation}
\label{attenuation}
\begin{aligned}
h_\ell(z_\ell) = \frac{P(z_{\ell})}{P(0)}=\text{exp}(-\sigma z_\ell),
\end{aligned}
\end{equation}
where ${z_{\ell}}$ is a propagation distance, ${P(z_\ell)}$ is a power level at ${z_\ell}$, and ${\sigma}$ is an attenuation coefficient determined by the visibility range.

%~\cite{visibility}

\subsubsection{Atmospheric Fading}
One of the conventional fading models of optical channels is a log-normal distribution, which usually stands for a stable atmospheric condition. The probability density function (PDF) of ${h_{a}}$ is given by
\begin{equation}
\label{lognormal}
\begin{aligned}
f_\text{LN}(h_a) = \frac{1}{2h_a \sqrt{2\pi \sigma_\text{X}^2}}\text{exp}\left(\frac{(\text{ln}h_a + 2\sigma_\text{X}^2 ) ^2}{8\sigma_\text{X}^2}\right),
\end{aligned}
\end{equation}
and ${\sigma_\text{X}}$ here, is approximated as ${\sigma_\text{R}/2}$, where ${\sigma_\text{R}^2}$ is the Rytov variance~\cite{channelbible}. 
On the other hand, $h_a$ follows gamma-gamma distribution considering scintillation:
\begin{equation}
\label{gammagamma}
\begin{aligned}
f_\text{GG}(h_a)=\frac{2(\alpha\beta)^{(\alpha + \beta)/2}}{\Gamma(\alpha)\Gamma(\beta)}h_a^{\frac{\alpha + \beta}{2} -1}K_{\alpha-\beta}\left(2\sqrt{\alpha\beta h_a}\right),
\end{aligned}
\end{equation}
where ${1/\alpha}$ and ${1/\beta}$ are the variances of the large-and small-scale eddies, and ${K_{\alpha-\beta}}$ is the modified Bessel function of the second kind~\cite{channelbible}.

\subsubsection{Pointing Error}

The two main factors determining the pointing error are beamwidth ${w_{z}}$ at a distance ${z}$ and beam displacement ${r}$~\cite{channel}. Coordinates of the UAV location are a superposition of the estimated point and Gaussian error. Then, the distribution of the beam displacement can be modeled as
\begin{equation}
\label{beamdisplacement}
\begin{aligned}
f_\text{dis}(r) = \frac{r}{\sigma_d^2}\text{exp}\left( -\frac{r^2}{2\sigma_d^2}\right),
\end{aligned}
\end{equation}
which is a Rayleigh distribution.
Here, ${\sigma_d}$ is the one-dimensional standard deviation of the Gaussian error, which satisfies ${\sigma_d = \sqrt{\sigma_\text{est}^2+\sigma_\text{jit}^2}}$ where ${\sigma_\text{est}}$ and ${\sigma_\text{jit}}$ are the standard deviation of the angle estimation error and the error induced by jitter, respectively.
Therefore, ${h_\text{p}}$ is given by ${r}$ as
\begin{equation}
\label{pointingerr}
\begin{aligned}
h_\text{p}(r,z)=A_\text{0}\text{exp}\left(-\frac{2r^2}{w_z^2}\right),
\end{aligned}
\end{equation}
as the beamwidth ${w_z}$ is approximately proportional to the distance and ${A_\text{0}}$ is the power gain at the center of the beam. In this case, the beamwidth can be described as ${w_z=z\cdot \theta_\text{div}}$ where ${\theta_\text{div}}$ is the beam divergence angle.

\section{Closed-form Angle Estimator}
\label{closedform}

If during the coarse-pointing stage, the uncertainty area of a UAV is wider than the typical beam-divergence angle, then a ground station scans the area so that the UAV can receive the beacon beam~\cite{rasterscan}. In practice, GPS information provides position information within an uncertainty range of a few meters. Therefore, we assume that the ground station selects a beam divergence angle to cover the entire uncertainty area when transmitting a beacon beam~\cite{2006spienews}. Given there is no scanning process using a beacon-beam footprint, this method is referred to as one-shot coarse pointing.

For successful link acquisition via one-shot coarse pointing, a UAV must detect a beacon beam transmitted from a ground station and transmit the beam back to the ground station. During this process, a beam-divergence angle must be sufficiently wide to completely cover the uncertainty area. Also, minimal beam divergence is required to intensify the beam and maximize the probability of detection. This trade-off demonstrates that the robustness of the link-acquisition step is directly affected by the uncertainty of the UAV location information. Here, the proposed advanced estimator reduces the outage probability of the coarse pointing process by improving the initial accuracy of UAV location information.

\begin{figure}[t]
	\begin{center}
		{\includegraphics[width=0.8\columnwidth,keepaspectratio]
			{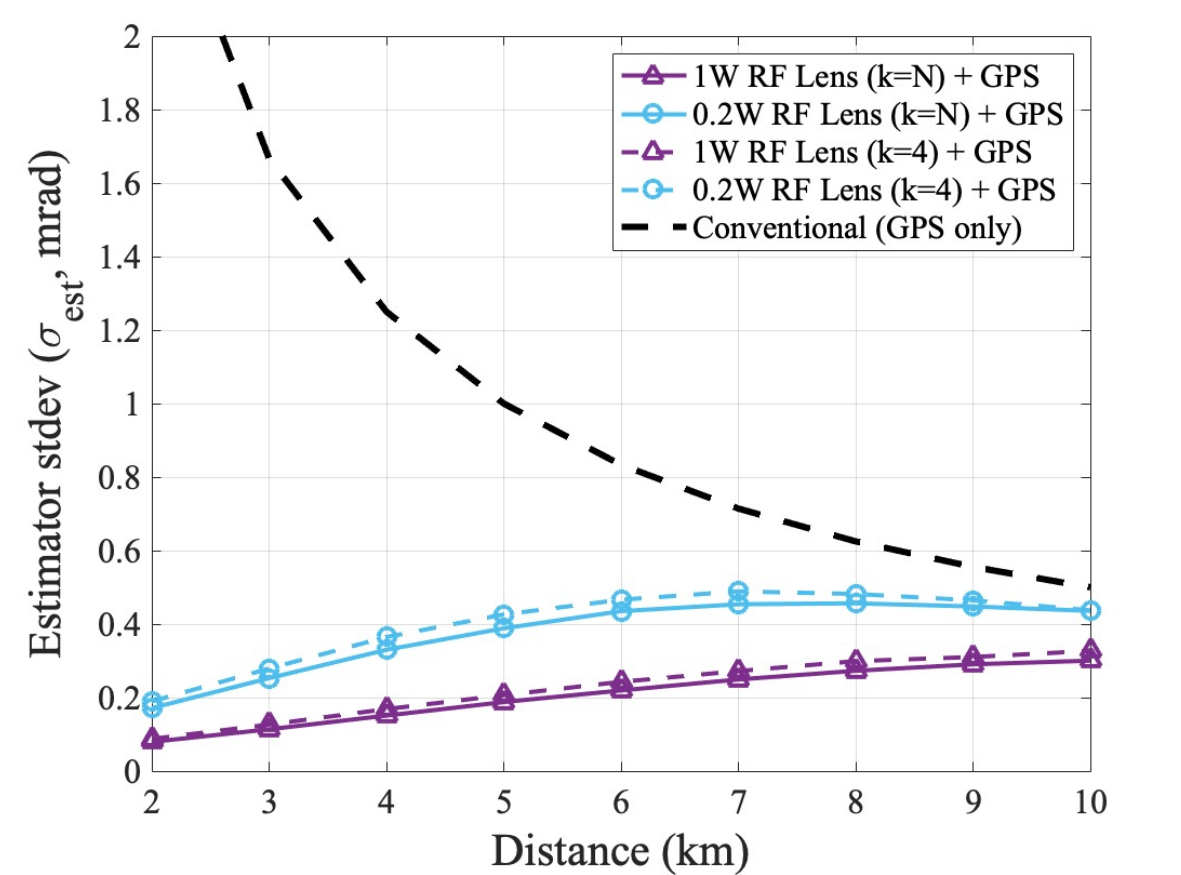}%
			\caption{Performance enhancement of the proposed estimator compared to the conventional GPS-only scenario.}
			\label{estimatorplot}
		}
	\end{center}
	\vspace{-15pt}
\end{figure}

We designed a highly accurate AoA estimation method with low complexity by fully utilizing both received RF signals and GPS information from a UAV. A closed-form suboptimal solution for the UAV location is first derived using the maximum a posteriori (MAP) criterion.
Assume that the antenna array receives a superposition of the signal vector ${\bold{y}}$ and a Gaussian noise vector~${\bold{n}}$. Then the conditional distribution of ${\bold{y}}$ given an AoA of ${\phi}$ can be written as
\begin{align}
\label{lensprob}
f_{\text{lens}}(\bold{y}|\phi)=\frac{1}{(\pi\sigma_{n})^{N}}\exp\left[-\frac{\norm{\bold{y}-g\bold{A}(\phi)\bold{s}e^{jb}}^{2}}{\sigma_{n}^{2}}\right]
\end{align} 
and the given GPS coordinate information can be equivalently converted into an angle ${\phi_{\text{gps}}}$. Thus the PDF of ${\phi}$ is given by
\begin{align}
\label{gpsprob}
f_{\text{gps}}(\phi)=\frac{1}{\sqrt{2\pi}\sigma_{\text{gps}}}\exp\left[-\frac{(\phi-\phi_{\text{gps}})^{2}}{2\sigma_{\text{gps}}^{2}}\right],
\end{align} 
where ${\sigma_{\text{gps}}}$ is the standard deviation of the GPS information and has a value of approximately 5 m divided by the distance to the UAV~\cite{2006spienews}.

\begin{algorithm}[t]
	\small
	\caption{Proposed coarse-pointing algorithm}
	\label{alg1}
	\begin{algorithmic}[1]
			\While {FSO link failure}
				\State ${u=[\frac{z}{d}\sin(\phi_{\text{gps}})-\frac{k}{2}+\frac{1+(-1)^N}{4}]+\frac{1-(-1)^N}{4}}$.
				\State ${v=[\frac{z}{d}\sin(\phi_{\text{gps}})+\frac{k}{2}+\frac{1+(-1)^N}{4}]+\frac{1-(-1)^N}{4}}$.
				\State Choose antenna indices $\forall i \in \mathbb{Z} : i \in [u, v]$ and connect to {\indent}each RF chain.
				\State Obtain ${\phi_\text{gps}, g, b,}$ and ${\sigma_\text{n}}$ during RF communication.
				\If{FSO channel is known}
					\State Choose beam divergence angle.
				\EndIf
				\State Compute (\ref{estimator}) and transmit beacon beam to UAV.
				\If{UAV detects beacon}
					\State Begin communication.
				\Else
					\State Compare ${t_0}$ and ${t_\text{rot}}$ to determine the appropriate policy.
				\EndIf
			\EndWhile
			\State \textbf{end}
	\end{algorithmic}
\end{algorithm}

\newtheorem{theorem}{Theorem}
\begin{theorem}{Given $\bold{y}$,  ${\phi_{\text{gps}}}$, ${g}$, ${b}$, and ${\sigma_\text{n}}$, a closed-form angle estimation can be expressed as follows
\footnote[1]{In practice, two-dimensional angle parameters are required for the complete location estimation of a UAV. Note that the result in (\ref{estimator}) is sufficient since the dimension extension of the antenna array can equivalently extend the dimension of estimation.}:
\begin{multline}
	\label{estimator}
	\phi_\text{est}=\phi_\text{gps}+\\
	\frac{e^{jb}\frac{\bold{y}}{g}^{*}\bold{A}'(\phi_\text{gps})\bold{s}+e^{-jb}\bold{s}^{*}\bold{A}'(\phi_\text{gps})\frac{\bold{y}}{g}-2\bold{s}^{*}\bold{A}(\phi_\text{gps})\bold{A}'(\phi_\text{gps})\bold{s}}{2\bold{s}^{*}\bold{A}'(\phi_\text{gps})^{2}\bold{s}+2\sigma_\text{n}^{2}/g^{2}\sigma_\text{gps}^{2}}.
\end{multline}
}
\label{theo}
\end{theorem}

\def\QEDmark{\ensuremath{\blacksquare}}
\def\proof{\emph{Proof: }}
\def\endproof{\hfill\QEDmark}

\proof
See the Appendix.
\endproof

\begin{comment}
\begin{figure*}[b]
\begin{center}
\line(1,0){516}
\begin{equation}
\begin{aligned}
\label{likediff}
	f'_{\text{lens}}({\bold{y}}|\phi)\approx-\frac{g}{\sigma_\text{n}^2}\big\{2g\bold{s}^*(\bold{A}(\phi_{\text{gps}})\bold{A}'(\phi_{\text{gps}})+\bold{A}'(\phi_{\text{gps}})^2\Delta\phi)\bold{s}-e^{jb}\bold{y}^*\bold{A}'(\phi_{\text{gps}})\bold{s}-e^{-jb}\bold{s}^*\bold{A}'(\phi_{\text{gps}})\bold{y}\big\}f_{\text{lens}}(\bold{y}|\phi)
\end{aligned}
\end{equation}
\begin{equation}
\begin{aligned}
	\label{finaleq}
	\Big[-\frac{g}{\sigma_\text{n}^2}\big\{2g\bold{s}^*(\bold{A}(\phi_{\text{gps}})\bold{A}'(\phi_{\text{gps}})&+\bold{A}'(\phi_{\text{gps}})^2\Delta\phi)\bold{s}-e^{jb}\bold{y}^*\bold{A}'(\phi_{\text{gps}})\bold{s}-e^{-jb}\bold{s}^*\bold{A}'(\phi_{\text{gps}})\bold{y}\big\}+\frac{\Delta\phi}{\sigma_{\text{gps}}^2}\Big]f_{\text{lens}}(\bold{y}|\phi)f_\text{gps}(\phi)=0
\end{aligned}
\end{equation}
\end{center}
\end{figure*}
\end{comment}

By Theorem~\ref{theo}, the ground station can accurately estimate the UAV location based on the received RF signal. Fig.~\ref{estimatorplot} demonstrates that, within a range of 10~km, the proposed estimator effectively reduces the standard deviation of UAV location inference. This results in a narrower uncertainty area because the uncertainty area is proportional to the square of the UAV location error. If the receiver connects four selected antennas to each RF chain, there is a slight performance degradation but a significant reduction in system complexity.

%Furthermore, the computation time of our estimator is more than a hundred times more brief than that of a brute-force search with a $100$~$\mu$rad resolution, which was tested on MATLAB.

%\section{Performance Analysis of the Proposed Algorithm}

\section{Proposed Coarse Pointing Algorithm}
\label{algorithm}
%\subsection{A Novel One-shot Coarse Pointing Algorithm}

Algorithm~\ref{alg1} is proposed; it allows flexible feeder-link acquisition for mobile UAVs supporting dynamic demands of data rate. Parameters for angle estimation are given through continuous RF tracking. To quickly connect the link, we now discuss optimizing the coarse-pointing factors: Beam divergence angle and coarse pointing policy. To this end, we assume that a ground station knows optical channel information.

%The proposed one-shot, coarse-pointing process for hybrid RF/FSO communications is summarized in Algorithm~\ref{alg1}. Continuous RF communication or tracking provides the parameters for angle estimation. If a ground station is aware of optical channel information, it is possible to select a beam divergence angle and coarse-pointing policy. A detailed description of the optimal selection of the beam divergence angle and policy is provided below.

%\subsection{Outage Probability}

\subsection{Beam Divergence Selection}
%Outage Probability
The coarse-pointing process can be successfully done when the received power ${P_{R}}$ is greater than the threshold power ${P_\text{th}}$. The outage probability ${P_\text{out}=\bold{Prob}[P_R<P_\text{th}]}$ is given by
\begin{equation}
\label{outageprob}
\begin{aligned}
P_\text{out} = \int_{0}^{\infty}f_\text{dis}(r)\bigg\{\int_{0}^{P_\text{th}}f \left(P_{R}|r\right)d{P_{R}}\bigg\}dr,
\end{aligned}
\end{equation}
where ${f(P_{R}|r)}$ is the PDF of ${P_{R}}$ for a given beam displacement ${r}$. The distribution is determined by using the channel model introduced in Section~\ref{chamod}. The numerical results in Section \ref{numres} allow the appropriate beam-divergence selection that minimizes the outage probability.

%\subsection{Processing Time}
\subsection{Coarse Pointing Policy}
%Processing Time
The processing time for coarse pointing strongly depends on the outage probability and channel-coherence time since the proposed algorithm must be executed until the UAV successfully receives a signal power greater than the threshold power. The channel-coherence time ${t_0}$ can be expressed as
\begin{equation}
\label{coherencetime}
\begin{aligned}
t_0=\bar{\rho}_0/v_\perp,
\end{aligned}
\end{equation}
where ${\bar{\rho}=\left(\frac{1}{2.91k^2C_n^2L}\right)^{3/5}}$ is the atmospheric correlation length, and ${v_\perp}$ is the wind speed perpendicular to the link~\cite{coherence}.

\begin{table}[t]
	\centering
	\caption{Simulation parameters}
	\small
	\label{tbl}
	\begin{tabular}{p{6cm} c}
	\hline
%	\bfseries{RF Lens Parameter} 				& \bfseries{Value} \\ 
%	\hline
%	RF transmitted power	& $1\,\,\si{\watt}$\\
%	Number of antenna elements ($N$)	& $41$\\
%	Lens diameter ($L$)	& $20\lambda$\\
%	Focal length ($f$)	& $20\lambda$\\
%	\hline
%	\hline
	\bfseries{Parameter} 				& \bfseries{Value} \\ 
	\hline
	Optical transmitted power ($P_\text{t}$)	& $1\,\,\si{\watt}$\\
	Optical threshold power ($P_\text{th}$)	& $1\,\,\si{\micro\watt}$\\
	Responsivity ($R$)	 & $0.5$\\ 
	Standard deviation of mechanical error ($\sigma_\text{jit}$)	& $3\,\,\si{\milli\radian}$\\
	Visibility range ($V$)	& $3,\,10\,\,\si{\kilo\metre}$\\
	Log-amplitude standard deviation ($\sigma_\text{X}$)	& $0.3$\\
	Gamma-gamma fading parameters ($\alpha, \beta$)	& $8.05, 1.03$\\
	Standard deviation of GPS error ($\sigma_\text{gps}$)	& $5\,\,\si{\meter}$\\
	\hline
	\end{tabular} 
	\vspace{-10pt}
\end{table}

We propose two different policies for attempting link acquisition. One re-estimates the AoA for every acquisition failure and transmits the beacon beam in a revised direction. The other relies on the initial AoA estimation result and waits until the UAV receives sufficient signal power as the channel fluctuates. Assuming that ${t_\text{rot}}$ is the time required to rotate a beam to the desired point, the average coarse pointing processing time ${\overline{T}_\text{re}}$ for the re-estimation policy is given by
\begin{equation}
\label{reestimation}
\begin{aligned}
\overline{T}_\text{re}=\frac{t_0+t_\text{rot}}{1-P_\text{out}}.
\end{aligned}
\end{equation}
In the single-estimation policy, the value of $r$ is fixed when the estimation is performed. The average processing time ${\overline{T}_\text{sing}}$ for the latter policy is
\begin{equation}
\label{singleestimation}
\begin{aligned}
\overline{T}_\text{sing}=\underset{r}{\mathbb{E}}\left[\frac{t_0}{1-F(r)}\right]+t_\text{rot},
\end{aligned}
\end{equation}
where ${F(r)=\int_{0}^{P_\text{th}}f \left(P_{R}|r\right)d{P_{R}}}$ is the outage probability for a given $r$.
If ${t_0{\ll}t_\text{rot}}$, then it is clear that the single-estimation strategy takes less time on average. On the other hand, ${t_0{\gg}t_\text{rot}}$ guarantees a shorter processing time for the re-estimation method. This can be proven as follows:
\begin{equation}
\label{timepf}
\begin{aligned}
\overline{T}_\text{re}=\frac{t_0}{1-P_\text{out}}=\frac{t_0}{1-\mathbb{E}\left[F(r)\right]}\leq\underset{r}{\mathbb{E}}\left[\frac{t_0}{1-F(r)}\right]=\overline{T}_\text{sing}.
\end{aligned}
\end{equation}
Since ${g(x)=\frac{1}{1-x}}$ is convex in $x\in[0,1)$, Jensen's inequality supports (\ref{timepf}). The ground station can select the best policy based on a comparison between $t_0$ and $t_\text{rot}$, which we demonstrate numerically in Section~\ref{numres}.

\section{Numerical Results}
\label{numres}

\begin{figure}[t]
	\begin{center}
		{\includegraphics[width=0.9\columnwidth,keepaspectratio]
			{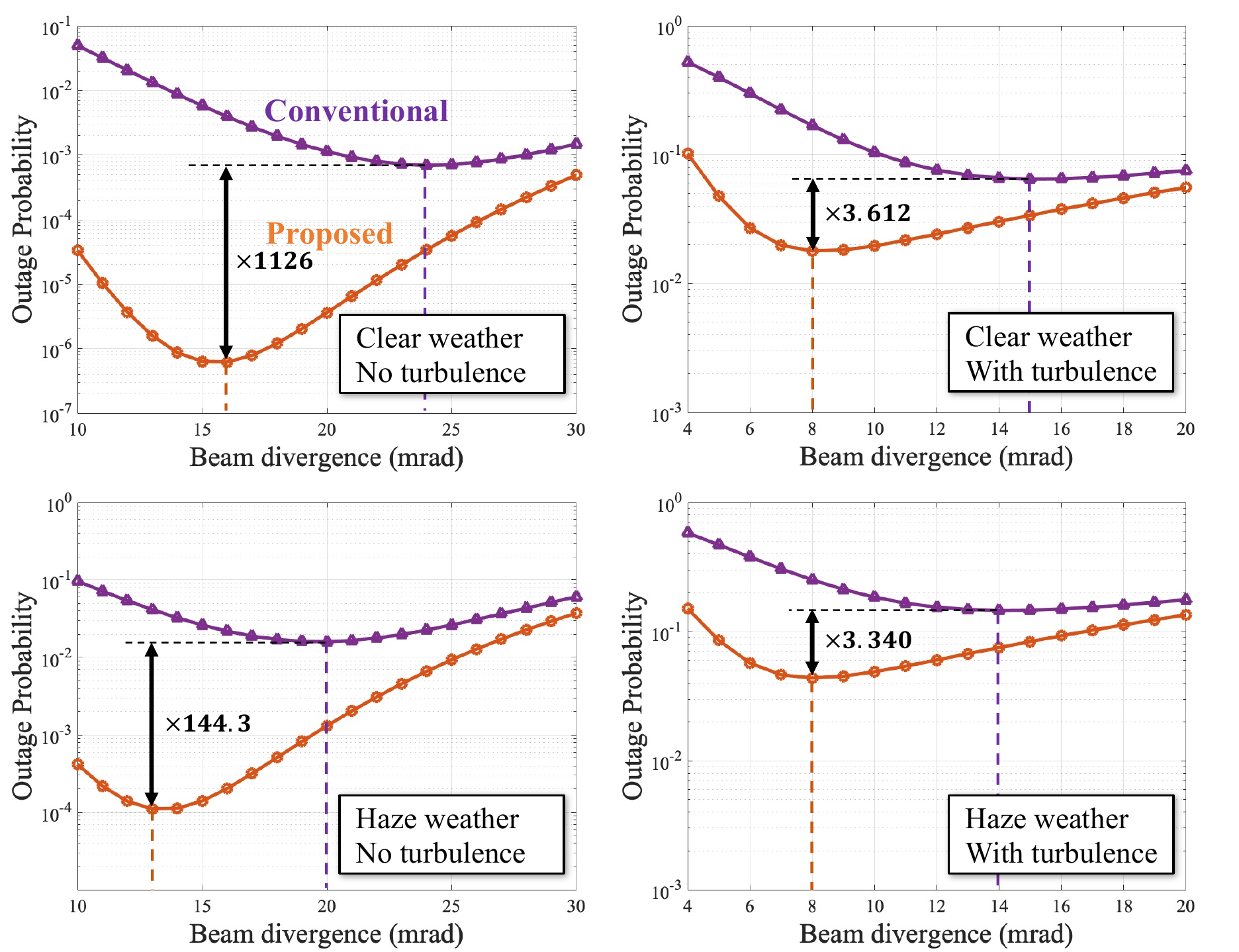}%
			\caption{Outage probability for a ${1}$~km link distance.}
			\label{outageplots}
		}
	\end{center}
	\vspace{-15pt}
\end{figure}
\begin{figure} 
    \centering
  \subfloat[\label{2a}]{%
       \includegraphics[height=0.5\linewidth]{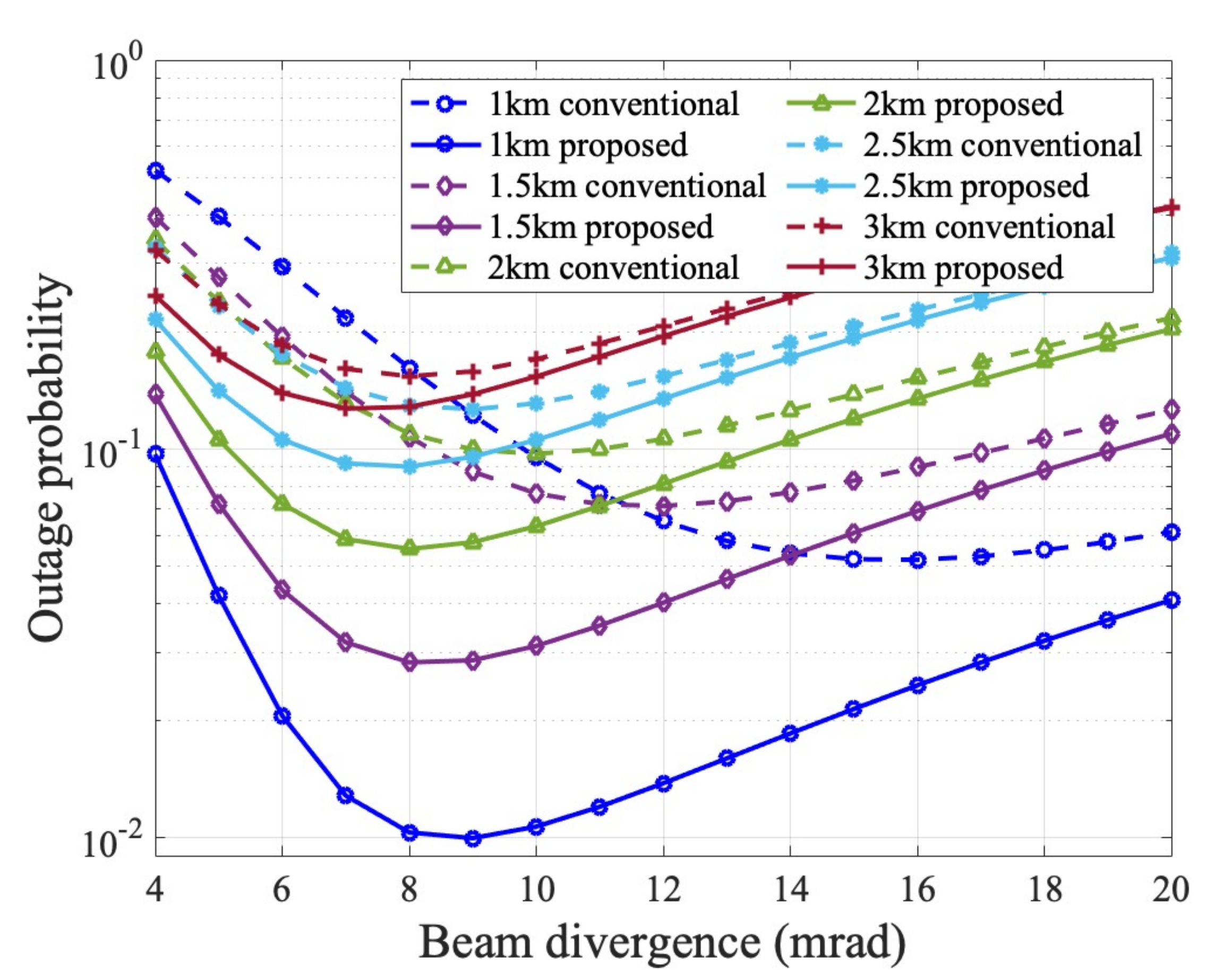}}
  \subfloat[\label{2b}]{%
        \includegraphics[height=0.5\linewidth]{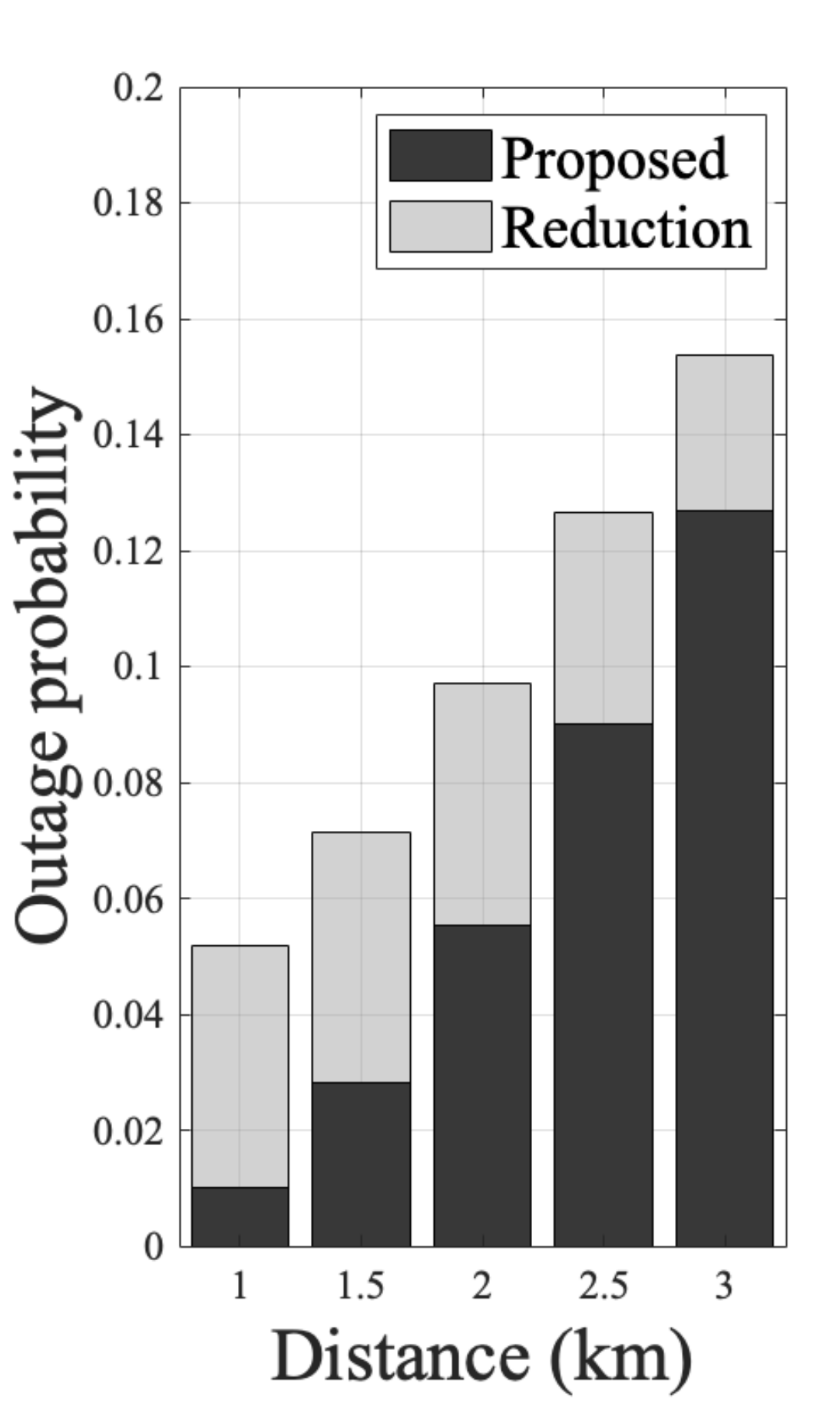}}
  \caption{(a) Outage probability under harsh weather for different distances and (b) the amount of reduction compared to the conventional method.}
  \label{distanceimage} 
  \vspace{-10pt}
\end{figure}

\begin{figure} 
    \centering
  \subfloat[\label{1a}]{%
       \includegraphics[width=0.7\linewidth]{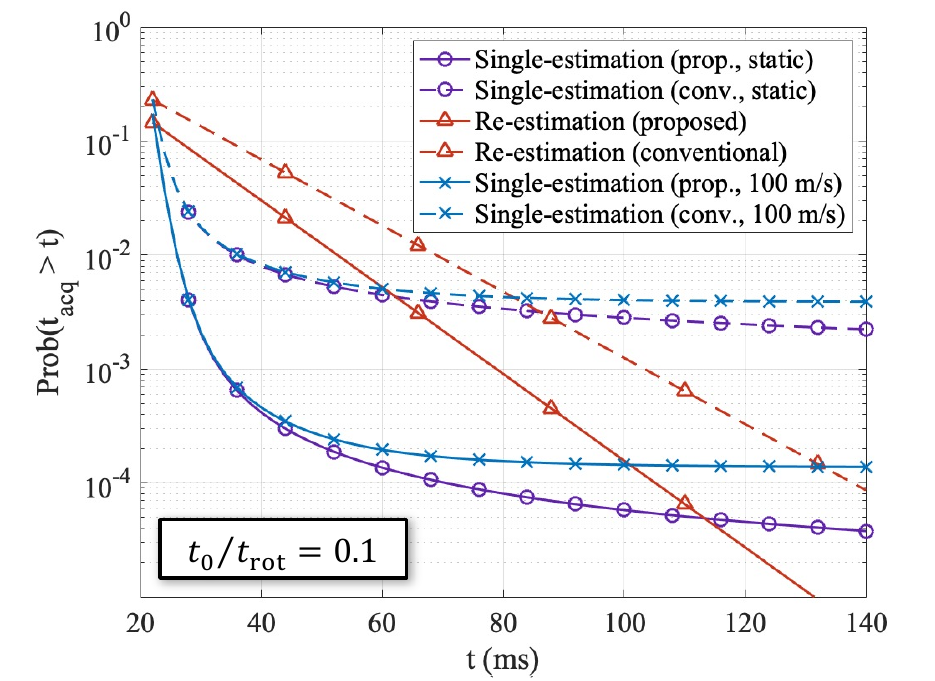}}
    \\
     \vspace{-10pt}
  \subfloat[\label{1b}]{%
        \includegraphics[width=0.7\linewidth]{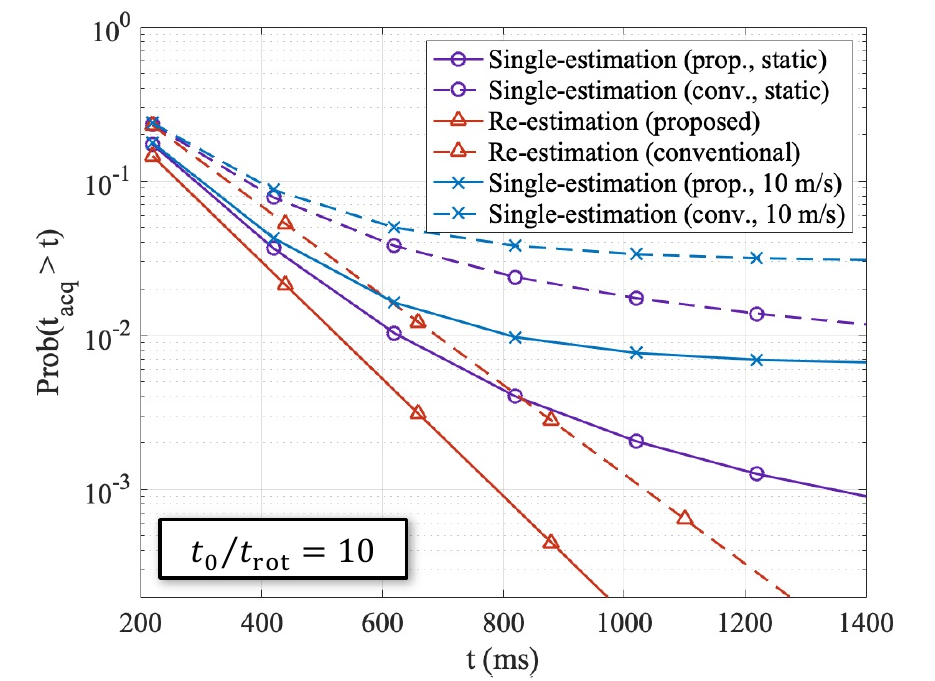}}
  \caption{Probability that a link is yet not connected at a time $t$ for (a) small and (b) large ${t_0}/{t_\text{rot}}$ values.}
  \label{cptime} 
  \vspace{-10pt}
\end{figure}

The detailed parameters used for our simulations are listed in Table~\ref{tbl}. In Fig.~\ref{outageplots}, the purple lines represent the outage probability when only GPS information is utilized for coarse pointing; the orange lines represent the proposed method. The lowest point on each line represents the lowest achievable outage probability and beam-divergence angle for the achievement. The result shows that the minimum outage probability is significantly reduced and can be achieved with a narrower beam divergence when using the proposed method. When the atmospheric channel is known to the ground station, lookup table-based, beam-divergence selection is possible.  Fig.~\ref{distanceimage} presents outage probability measurements with varying distances between the ground station and UAV. As the distance increases, the gap narrows between the solid and dashed lines. This shows that when the link distance is short, the proposed method is more effective. This is because the intensity of an RF signal attenuates significantly with propagation distance.

Fig.~\ref{cptime} shows the numerical results for the processing time of one-shot, coarse pointing using (\ref{reestimation}) and (\ref{singleestimation}). The values indicate the probability that the link-acquisition time ${t_\text{acq}}$ is greater than $t$. The value of ${t_\text{rot}}$ is fixed at $20$~ms and the link distance is assumed to be $2$~km. As shown in Fig.~\ref{cptime}, when ${t_0}$ is much smaller than ${t_\text{rot}}$, the processing time of the single-estimation method is almost certainly faster than that of the re-estimation method. Conversely, the re-estimation method is always faster when ${t_0}$ is sufficiently large.
When mobility exists, the performance degrades over time only for a single-estimation policy since the estimated AoA is not updated while the UAV is moving.

\section{Conclusion}
\label{conclusion}

In this letter, we derived, as a component of a PAT system, a closed-form solution for angle estimation and proposed a fast one-shot coarse-pointing algorithm for hybrid RF/FSO communication systems. We showed that the proposed algorithm effectively reduces the outage probability of link acquisition. Additionally, when an accurate prediction of the optical channel conditions is available, it is possible to find the beam-divergence angle that minimizes the outage probability. Further, we formulated and analyzed the processing time of the one-shot, coarse-pointing process. Finally, we numerically verified that, in terms of outage probability and processing time, it is superior to the conventional method.
\appendix[Proof of Theorem~\ref{theo}]
Following MAP criterion, the proposed estimator can be formulated as
\begin{equation}
\label{lensproblem}
\begin{aligned}
\phi_\text{est}=\underset{\phi}{\argmax}f(\phi|\bold{y})=\underset{\phi}{\argmax}\frac{f_\text{lens}(\bold{y}|\phi)f_\text{gps}(\phi)}{f(\bold{y})},
\end{aligned}
\end{equation}
where ${f(\bold{y})=\int f_\text{lens}(\bold{y}|\phi)f_\text{gps}(\phi)\,d\phi}$.
Given $\bold{y}$, since ${f(\bold{y})}$ is a constant, the solution of (\ref{lensproblem}) can be obtained by searching for the local maximum of ${f_\text{lens}(\bold{y}|\phi)f_\text{gps}(\phi)}$ with respect to $\phi$ if it is a concave function. To this end, we apply the first-order Taylor approximation ${\bold{A}(\phi)=\bold{A}(\phi_{\text{gps}})+\bold{A}'(\phi_{\text{gps}})(\phi-\phi_{\text{gps}})}$ to the following function:
\begin{equation}
\begin{aligned}
\label{lensprobdiff}
f'_{\text{lens}}(\bold{y}|\phi)=&\left\{\frac{\partial}{\partial\phi}{\norm{\bold{y}-g\bold{A}(\phi){\bold{s}}e^{jb}}^{2}}\right\}\\&\cdot\frac{1}{(\pi\sigma_{n})^{N}}\exp\left[-\frac{\norm{\bold{y}-g\bold{A}(\phi){\bold{s}}e^{jb}}^{2}}{\sigma_{n}^{2}}\right].
\end{aligned} 
\end{equation}
Note that a diagonal matrix $\bold{A}'(\phi) = \bold{H}'(\phi)[u:v,u:v]$ and $\bold{H}'(\phi) = \frac{d}{d\phi}\bold{H}(\phi)$. Each component of $\bold{H}'(\phi)$ then satisfies $\bold{H}'(\phi)_{n+\frac{N+1}{2},n+\frac{N+1}{2}} = \frac{L^{2}\cos(\phi)}{\lambda\sqrt{z}{{\phi_{1}}^2}}\{\sin(\phi_{1})-\phi_{1}\cos({\phi_{1}})\}$, where $\phi_{1} = \frac{L}{\lambda}(\frac{dn}{z}-\sin({\phi}))$ and $n$ is a set from $-(N-1)/2$ to $(N-1)/2$. Hence, the derivatives of ${f_{\text{gps}}(\phi)}$ and ${f_{\text{lens}}({\bold{y}}|\phi)}$ are given by
\begin{equation}
\begin{aligned}
	\label{gpsdiff}
	f'_\text{gps}(\phi)=\frac{\Delta\phi}{\sigma_{\text{gps}}^2}f_\text{gps}(\phi),\hspace{12.2em}
\end{aligned}
\end{equation}
\begin{equation}
\begin{aligned}
\label{likediff}
	f'_{\text{lens}}({\bold{y}}|\phi)\approx&-\frac{g}{\sigma_\text{n}^2}\big\{2g\bold{s}^*(\bold{A}(\phi_{\text{gps}})\bold{A}'(\phi_{\text{gps}})+\bold{A}'(\phi_{\text{gps}})^2\Delta\phi)\bold{s}\\&-e^{jb}\bold{y}^*\bold{A}'(\phi_{\text{gps}})\bold{s}-e^{-jb}\bold{s}^*\bold{A}'(\phi_{\text{gps}})\bold{y}\big\}f_{\text{lens}}(\bold{y}|\phi),
\end{aligned}
\end{equation}
respectively, where ${\Delta\phi=\phi-\phi_{\text{gps}}}$. Therefore, by solving ${\frac{\partial}{\partial\phi} f_\text{lens}(\bold{y}|\phi)f_\text{gps}(\phi)=0}$, the local maximum $\phi_\text{est}$ satisfies
\begin{equation}
\begin{aligned}
	\label{finaleq}
	\Big[-\frac{g}{\sigma_\text{n}^2}\big\{2g\bold{s}^*(\bold{A}(\phi_{\text{gps}})\bold{A}'(\phi_{\text{gps}})+\bold{A}'(\phi_{\text{gps}})^2\Delta\phi)\bold{s}&
\\-e^{jb}\bold{y}^*\bold{A}'(\phi_{\text{gps}})\bold{s}-e^{-jb}\bold{s}^*\bold{A}'(\phi_{\text{gps}})\bold{y}\big\}+\frac{\Delta\phi}{\sigma_{\text{gps}}^2}\Big]&
\\{\cdot}f_{\text{lens}}(\bold{y}|\phi)f_\text{gps}(\phi)&=0,
\end{aligned}
\end{equation}
which has the unique solution of (\ref{estimator}).

\bibliographystyle{IEEEtran}
\bibliography{FSO_WCL}

\end{document}